# Structure Preserving Transformations for Practical Model-based Systems Engineering

Siyuan Ji, Member IEEE
*Department of Computer Science*
*the Univeristy of York*
York, UK
siyuan.ji@york.ac.uk

Michael Wilkinson
*BAE Systems Maritime*
Barrow-in-Furness, UK
michael.wilkinson1@baesystems.com

C.E. Dickerson, Senior Member IEEE
*School of Mechanical, Electrical and*
*Manufacturing Engineering*
*Loughborough University*
Loughborough, UK
c.dickerson@lboro.ac.uk

*Abstract*—In this third decade of systems engineering in the twenty-first century, it is important to develop and demonstrate practical methods to exploit machine-readable models in the engineering of systems. Substantial investment has been made in languages and modelling tools for developing models. A key problem is that system architects and engineers work in a multidisciplinary environment in which models are not the product of any one individual. This paper provides preliminary results of a formal approach to specify models and structure preserving transformations between them that support model synchronization. This is an important area of research and practice in software engineering. However, it is limited to synchronization at the code level of systems. This paper leverages previous research of the authors to define a core fractal for interpretation of concepts into model specifications and transformation between models. This fractal is used to extend the concept of synchronization of models to the system level and is demonstrated through a practical engineering example for an advanced driver assistance system.

## I. Introduction

During the first two decades of the twentieth century the systems engineering community has seen significant advances in Model-based Systems Engineering (MBSE) and commercially available tools to support its practice. This is evident from the INCOSE Systems Engineering Handbook [1] and Vision 2020, and the development of the OMG Systems Modeling Language (SysML) and standards for modeling tools [2]. However, most existing methodologies such as the Object Oriented Systems Engineering Methodology (OOSEM) [3] are defined informally, so there is potential for ontological inconsistencies between them and with formal approaches such as the Object-Process Methodology (OPM) [4].

Furthermore, despite these expansive advances, it has been argued whether the advances are reflected by the uptake in the practice of MBSE. In the recent 2022 CSER conference [5], Wilkinson argued in a keynote address that the benefits of system architecting and modeling (such as re-use) are difficult to evidence and realize. In the 2020 survey of post graduate engineering students at MIT [6], only a small fraction of the companies reported using SysML or OPM. In the 2015 INCOSE survey [7], MBSE practitioners predominantly did Architecture and Requirements Definition. Dickerson and Ji noted in [8] that evidence similar to the 2015 findings can be traced back as early as the late 1990s in the commercial practice of systems engineering. It should be no surprise then that a key feature of the current system architecting tools as standardized by the OMG is the traceability of requirements in machine-readable system models.

The need for further such features of machine-readable system models is then clear if the adoption of MBSE is to gain momentum. One such feature is the synchronization of models. This is an active area of research in software engineering but is focused primarily at the implementation level. The meaning of the term synchronization in this context is that multiple programmers are working on various parts of the computer code and indeed sometimes on the same part. The analogy for systems engineering is clear. The multiple disciplines and system component engineers will be working on various system models that are interrelated and sometimes are the same model. A higher level of abstraction is needed.

Basic operations of computer coding such as create, read, update, and delete (CRUD) will be abstracted to the system level. The graphical models of SysML will be represented by matrices. Complete automation of model synchronization in MBSE is not practical due to system complexity and the human interpretation of domain knowledge into system models. Instead, what is needed is a tool with mathematically based algorithms that run in the background and work with engineers (rather than replacing them). Tools like this are referred to as joint cognitive systems.

The approach taken in this paper introduces a core formalism that structures the processes of interpretation and model transformation that underlie MBSE. As such it is an architecture (of MBSE) in the abstract sense of the term defined by the authors in [9]. This is a key contribution. Other significant contributions are:

- Structure preserving transformations for practical MBSE that underlie the core formalism
- Specification of models in a synchronized manner to achieve desired model traceability that is machine readable
- Synchronization of changes in ways agnostic of where the changes are introduced
- Demonstration through an ADAS case study

The formalism is implementable in a joint cognitive paradigm: (i) automating the specification and synchronization of models using formal transformations where possible and (ii) supporting system architects and designers to fill in the gaps.

The preliminary results offered in this paper also show that MBSE cannot be practical for engineering purposes without model synchronization. In fact, without this capability, MBSE is at risk of being where IT was with systems in the 1990s, namely a state of semantic confusion referred to as a Tower of Babel.

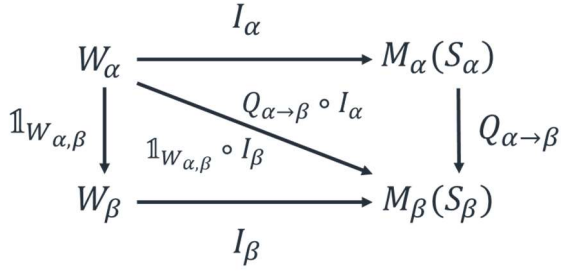

Fig. 1. The core of the formalisim for MBSE.

## II. THE CORE OF THE FORMALISM

The aim of this work is to propose a formalism that enables the development and specification of models in a consistent way that would in turn enable the synchronization of the models when changes to the models are introduced. In this section, we define the core elements of the formalism and how these elements are related, as depicted in Figure 1.

### A. Domain Knowledge

**Domain knowledge** – Denoted by $W$, Domain Knowledge is domain-specific information that captures what engineers know about that domain. Its content is, in principle, unbounded. However, only knowledge important to solving the engineering problem at hand will be used to develop models. Domain knowledge is usually represented in natural language, e.g., in well-formed and structured sentences when concerning system requirements.

### B. Structures

**Semantic Structure** – A Semantic Structure, $S$, is a structure constructed by using a modelling language in which the semantics of the language is well defined. Such a structure is also regarded as a 'clean' model in which the structure is not populated by any domain knowledge. Examples of Semantic Structures are depicted in Figure 2, where a clean Use Case diagram and Activity diagram are drawn. The structures embodied in these diagrams conform to the metamodel of the UML/SysML but do not provide any domain-specific meaning.

**Domain Structure** – When a Semantic Structure is populated by domain knowledge, it becomes a Domain Structure, $M$. A Domain Structure is also regarded as a populated model. An example of a Domain Structure can be found in Figure 3, where the populated Use Case diagram conveys domain-specific meanings.

### C. Structuring Preserving Transformations

**Semantic Transformation** – A Semantic Transformation, $Q_{\alpha \to \beta}$, is a transformation that transforms a Semantic Structure, $S_\alpha$ into another Semantic Structure, $S_\beta$, i.e., $Q_{\alpha \to \beta}: S_\alpha \to S_\beta$. The two structures are also notionally regarded respectively as a pair of source and target structures when discussed in the context of $Q_{\alpha \to \beta}$. Considering Semantic Structures as (clean) models, $Q_{\alpha \to \beta}$ can be considered as a model transformation, which have been well-studied in Model-Driven (Software) Development [10].

**Interpretation** – An interpretation, $I$, interprets the domain knowledge, as described in $W$, into a Semantic Structure, $S$, such that the Semantic Structure will be fully populated to become a Domain Structure, $M$. The resultant

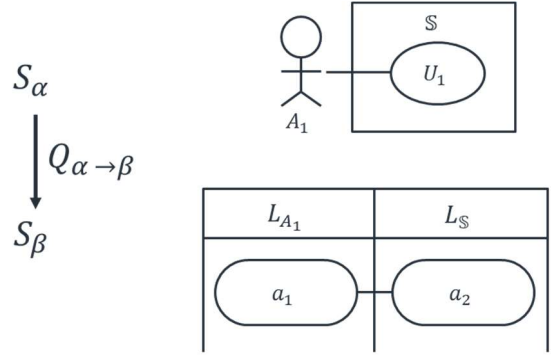

Figure 2. Definition of Semantic Transformations between Use Case structures, $S_\alpha$ and Activity structures, $S_\beta$.

Domain Structure, i.e., the populated model, represents a subset of the information content in $W$.

Despite the direction of $I_\alpha$ going from Domain Knowledge to the Semantic Structure, the underlying construct implies a modelling approach where knowledge is retrieved based on the semantic meanings embodied in the Semantic Structures. This is consistent with well-known Theories of Design. For example, the so-called C-K Theory developed by Hatchuel and Benoît [11] explains how knowledge is used to stimulate new concepts and to validate them.

### D. Composition of Transformations

The final elements in the core formalism are the **Compositions of Transformations**, depicted by the diagonal lines in Figure 1. Instead of directly transforming Domain Knowledge, $W_\alpha$, into the Semantic Structure, $S_\beta$, this directed arrow utilizes the concept of composition as in Category Theory to form two-steps transformations to ensure consistency between the two models, $M_\alpha$ and $M_\beta$. Specifically, there are two compositions:

- $Q_{\alpha \to \beta} \circ I_\alpha$, a composition that simultaneously preserves the semantics in $W_a$ and the structure of $S_a$. This composition is used to develop an initial $M_\beta$ that is consistent with $M_\alpha$. Details of how this composition is executed will be explained in Section III-C.

- $\mathbb{1}_{W_{\alpha,\beta}} \circ I_\beta$, a composition that preserves the interrelationship, $\mathbb{1}_{W_{\alpha,\beta}}$, between $W_a$ and $W_\beta$ into $M_\beta$ through $I_\beta$. This composition is used to verify the consistency between $M_\alpha$ and $M_\beta$. Details of how this composition is executed will be explained in Section III-F.

## III. FRACTAL PROCESSES

In this section, the formalism introduced in the previous section will be broken down to a series of fractal processes to illustrate how the formalism enables the development of consistent models. To further illustrate how the processes are applied in practice, we will be applying the processes to the architecting and modelling of an Emission Control Governor (ECG) system [8] to control engine emissions through engine calibration tuning while the vehicle motion is controlled by an Advanced Driver Assistance System (ADAS). In simple terms, the ECG would take motion control commands from

ADAS and determine optimal and robust engine calibration points for the engine to calibrate. Such a system can be achieved through coordinated control architecture such as the one presented by Lin [12]. Due to space limitation, we will only focus on a simple operational concept: *ECG determines engine torque value for an ADSA torque demand*. We claim that despite the usage of this simple case study, the fractal processes are agnostic to the domain of application and the overarching MBSE methodology adopted.

*A. Step 1: Define Semantic Transformation, $Q_{\alpha \to \beta}$*

The processes start with defining or selecting the types of the Semantic Structures, $S_\alpha$ and $S_\beta$, such that a directed Semantic Transformation between the two structures, $Q_{\alpha \to \beta}$ can be established. Such a transformation can be formally defined using the formalism as described in Ref. [13] for any generic rule-based model transformation and implemented using model transformation languages such as the EMF Henshin language built based on the Triple Graph Grammar paradigm [14] or the Epsilon transformation language [15].

In the context of model development and specification for MBSE, the rules of the Semantic Transformations should be constructed based on the semantic dependencies between the two structures. For example, Semantic Transformation from a Use Case diagram to an Activity diagram would utilize the *semantic dependencies* as derived from their metamodels. According to UML 2.5.1 [16], the semantic meaning of a Use Case when applied to a subject, is described as specifying "*a set of behaviors performed by that subject…*"; while the semantic meaning of an Action is defined precisely as "*a fundamental unit of executable functionality contained, directly or indirectly, within a Behavior*". The dependency between a Use Case and Actions thus becomes evident through the shared usage of *Behavior*. This dependency can be formulated as follows: *a Use Case can be realized by a set of Actions*. These derived semantic dependencies are also referred to as *cross-model dependencies* that properly connect the models to establish traceability and consistency between the models.

For practical MBSE, the definition of $Q_{\alpha \to \beta}$ should be agonistic to the model transformation paradigm and language [8] used. Engineers could picture $Q_{\alpha \to \beta}$ as a 'template' for how $S_\beta$ would look with a given $S_\alpha$. It is important to acknowledge that such a template would primarily constrain $S_\beta$, but not entirely on $M_\beta$ such that there is a degree of flexibility when it comes to filling the template, i.e., Interpretation of $W_\alpha$ into $S_\beta$ that is explained in the subsequent subsection. For instance, as depicted in Figure 2, the following transformation rules are executed and illustrated (note, for convenience, letter-based notations are introduced as the identifiers for the corresponding model element):

- The System (denoted by $\mathbb{S}$) and each Actor (denoted by $A_i$) in the Use Case diagram is transformed into a corresponding Swimlane (denoted by $L_i$) in the Activity diagram: e.g., $\mathbb{S} \to L_\mathbb{S}$ and $A_1 \to L_{A_1}$.

- Each Use Case in the Use Case diagram is transformed into a set of two Actions (as a starting point), with one action allocated to the System Swimlane and the other action allocated to the Actor Swimlane corresponding to the Actor in which the Use Case was associated to in the Use Case diagram, for example, $U_1 \to (a_1, a_2)$ with Allocate($a_1, SL_{A_1}$) and Allocate($a_2, SL_{Sys}$).

- The two Actions are further connected by a precedence relation in which its direction of flow is yet undecided. This connection preserves the association relationship between $A_1$ and $U_1$, Associate($A_1, U_1$).

It is worth noting that following a proper binary transformation [17] that preserves Associate($A_1, U_1$), one would derive a dependency between $a_2$ and $L_{A_1}$. However, modelling this dependency directly in the Activity diagram would violate the syntax rules constrained by the metamodel of the Activity diagram. This motivated us to formulate a rule that also creates an Action $a_1$ to allow structure to be preserved. Further, the actual number of Actions that $U_1$ transforms into cannot be determined at this stage, as this is subject to the Domain Knowledge to be interpreted into the two models in the later steps. Therefore, one observes naturally that such transformations only semi-automate the creation of the structure, leaving the architect flexibility to design the exact details when completing the model; hence the transformation being joint cognitive.

*B. Step 2: Specify Model, $M_\alpha$*

The second step in the processes concerns the specification of $M_\alpha$ through the interpretation of the domain knowledge as captured in $W_\alpha$. Thus, the step also involves the definition of $I_\alpha$.

The definition of $I_\alpha$ should be partially guided by the Semantic Structure, $S_\alpha$, which the Domain Knowledge will be interpreted into and partially by the intended domain-specific usage of $S_\alpha$. For example, in addition to how a Use Case is used to model behaviors of the system, a particular MBSE approach may further restrict a Use Case diagram to be only used at a particular level, e.g., often at the highest level of abstraction in the system hierarchy, to model functionalities that the system provides. As such, this would restrict the knowledge that one would need to retrieve, which has a determining effect on how $I_\alpha$ should be defined in a structure preserving way.

Following the above example, let us consider an MBSE approach where a Use Case diagram is used to capture functional requirements at the system level. Specifically, the structure in which a Use Case is associated with an Actor, abbreviated as a Use Case-Association-Actor structure (U-A-A Structure), is used to only model the active provision of a function by the System [8]. Following this specific modelling constraint, one would need to retrieve Domain Knowledge that is expressed in a similar way, describing how the system provides a function to its users in an active way. For the ECG system, one derives two functional requirements that exhibit such a pattern:

- R1: ECG shall receive torque demand from ADAS

- R2: ECG shall govern engine torque

Then, $W_\alpha$ is said to contain two sentences, R1 and R2. These two sentences can then be interpreted into the U-A-A Structure with the following rules that form $I_\alpha$:

- Identify the System in $W_\alpha$, which is then interpreted into the «subsystem», representing the subject of interest (system); in this case: ECG.

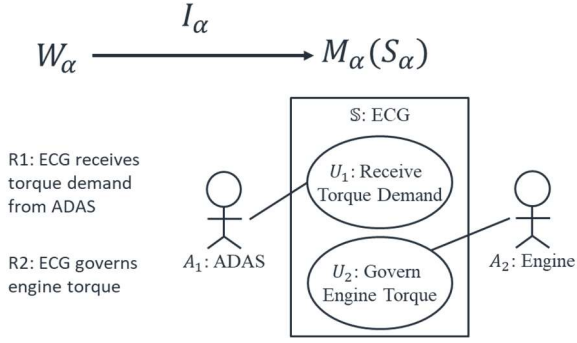

Figure 3. Specification of $M_\alpha$ through interpreting $W_\alpha$ into the Semantic Structure, $S_\alpha$ of $M_\alpha$ following the rules of $I_\alpha$

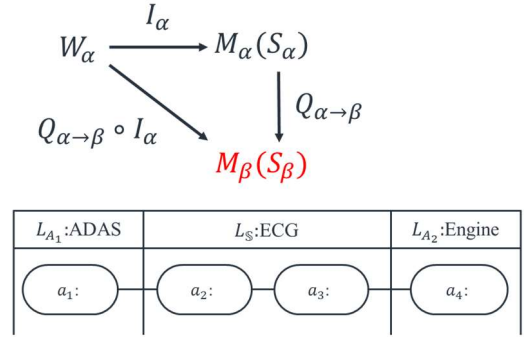

Figure 4. Initial specifcation of $M_\beta$ based on the composition of the transformations, $Q_{\alpha\to\beta} \circ I_\alpha$.

- For each sentence in $W_\alpha$, the *object* derives an element in the environment of the *subject* (system), which is interpreted into an Actor:
    - R1 derives an 'ADAS' Actor
    - R2 derives an 'Engine' Actor
- The *verb* phrase in each sentence is interpreted into a Use Case, following a verb-noun construct:
    - R1 derives an 'Receive Torque Demand' Use Case
    - R2 derives an 'Govern Engine Torque' Use Case
- An Association is used to connect the derived Actor and the Use Case with respect to each sentence:
    - 'ADAS' Actor is associated with 'Receive Torque Demand' Use Case
    - 'Engine' Actor is associated with 'Govern Engine Torque' Use Case

The above rules are generic in that it is agnostic to domains of interest. However, not all functional requirements can be phrased in a way that naturally follows a simple *subject-verb-object* structure. Therefore, to accommodate the usage of natural languages, a complete $I_\alpha$ would be expanded to address different sentence structures that would also adhere to an interpretation into the U-A-A Structure in the Use Case diagram. This step and the resultant Use Case diagram derived from the application of $I_\alpha$ to $W_\alpha$ is depicted in Figure 3.

### C. Step 3: Perform Composition, $Q_{\alpha\to\beta} \circ I_\alpha$

The third step in the processes is to derive an initial $M_\beta$, through a composition of the two transformations defined in the previous step, i.e., $Q_{\alpha\to\beta} \circ I_\alpha$. This step can be intuitively understood as a continued interpretation of $I_\alpha$ into $M_\beta$ that conforms to the structure, $S_\beta$, transformed through $Q_{\alpha\to\beta}$. For instance, as shown in Figure 4, following the previous two steps, the three Swimlanes in the initial Activity diagram can now be automatically named to maintain the consistency between the two models. The pre-assigned identifiers, e.g., $A_1$ and $L_{A_1}$, facilitate the automation of this step.

There are parts of $W_\alpha$ that are not preserved into $M_\beta$ through $Q_{\alpha\to\beta} \circ I_\alpha$. This is expected in general as the graphical models are meant to present different perspectives on the system, i.e., the perspective of $M_\beta$ is only partially aligned with $W_\alpha$. The composition, $Q_{\alpha\to\beta} \circ I_\alpha$ reveals the alignment.

### D. Step 4: Specify Target Model, $M_\beta$

To complete the specification of $M_\beta$, we next retrieve additional Domain Knowledge to fill in the remaining unpopulated structure of $M_\beta$ through an Interpretation, $I_\beta$. In the running case study, as depicted in Figure 4, this step would require system engineers to identify knowledge that would allow the specification of:

- the set of Actions ($a_1, a_2, a_3, a_4$); and
- the functional flows between the Actions.

To demonstrate how $I_\beta$ can be constructed and the knowledge required to build $W_\beta$, we assume that upon a stakeholder meeting with domain engineers, the two initial functional requirements are refined into a set of two new requirements that are specified using a specific structured natural language technique, EARS [18]:

- R1': When ADAS makes a torque demand, the ECG shall receive this torque demand from ADAS.
- R2': When ECG receives torque demand from ADAS, it shall determine an Engine torque for the engine to calibrate against.

The usage of EARS helps establishing the precedence relations between requirements through identifying pre- and post- conditions. In this case, it is apparent that the natural order of occurrence is ADAS makes a torque demand → ECG receives this demand → ECG determines Engine torque → Engine calibrates against the determined value. This becomes the basis for defining the set of actions, as shown in Figure 5.

For simplicity, we have intentionally defined only two Actions for each of the refined requirements, which is the minimum number to reflect a 'communication' between the Actor and the System (modelled as Swimlanes in $M_\beta$). It could well be that when it comes to interface definition at a later stage [8], one of the two Actions is no longer applicable, e.g., ECG does not have an active receiver for the ADAS command, but passively transforms a digital torque demand from ADAS into an Engine torque calibration value. Such changes to the model can be dealt with by synchronization of model changes as illustrated in Section V and does not violate the cross-model dependencies established so far.

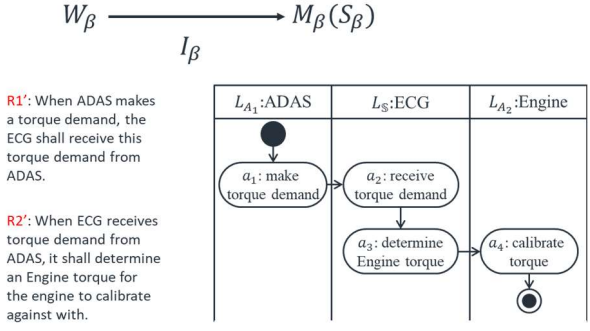

Figure 5. Complete specification of $M_\beta$ through interpretation of additional and refined knowledge.

In practice, one could also derive, from the requirements, a much more complex set of Actions that may be connected in series, in parallel or both. Despite the complexity, it is worth noting that rules of $Q_{\alpha \to \beta}$ (a Use Case is transformed into a set of Actions) specified in Section III-A is not violated. As a joint cognitive system, the structure of the set of Actions requires input from the architect.

Step 4 marks the completion of model specification because both $M_\alpha$ and $M_\beta$ are now specified with details and their structure and semantics are consistent with each other. At this point, it is observed that cross-model dependencies between two models are achieved at two levels:

*1)* At the structure level – model elements (entities and relations) of the two models are linked based on their semantic constructs as directed by their metamodels and intended usage in practice.

*2)* At the knowledge level – while constrained by the links established at the structure level, cross-model dependencies are further refined and reinforced by the knowledge interpreted into the structures.

### E. Step 5: Derive Domain Dependency, $\mathbb{1}_{W_{\alpha,\beta}}$

Following the previous step, the next two steps aim to verify the correctness of $M_\beta$ and consistency between the two models, $M_\alpha$ and $M_\beta$ through exploring the relationships between the two sets of Domain Knowledge, $W_\alpha$ and $W_\beta$ (Step 5), and interpretation of these dependencies into $M_\beta$ via $I_\beta$ (Step 6). Essentially, the idea is to use the bottom-left half of the formalism to verify the results achieved through the top-right half of the formalism.

Using the results obtained from the previous steps in the ECG case study, it is clear that $W_\beta$ is a one-to-one elaboration of $W_\alpha$ where each sentence in $W_\alpha$ is elaborated into further details to allow the interpretation of functional flow and functional allocation. As such, the relationship between them can be expressed by $W_\alpha \subset W_\beta$, i.e., all information contained in $W_\alpha$ is also represented in $W_\beta$, but not the other way around.

This relationship is anticipated as the case study is about transforming a Use Case diagram to an Activity diagram, representing a process of functional elaboration that is standard in many systems engineering processes. In general, there could be different types of dependencies between $W_\alpha$ and $W_\beta$, representing different engineering processes or paradigms, for instance, reversing engineering and abstraction where $W_\beta \subset W_\alpha$, e.g., abstracting an Activity diagram into a higher-level Use Case diagram.

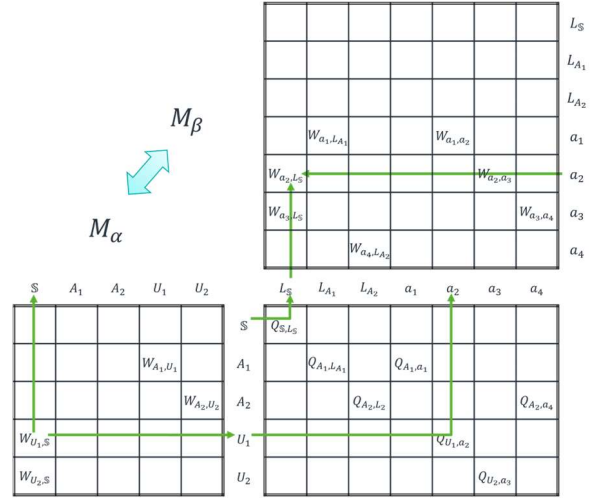

Figure 6. Synchrnoisation of source and target models using the ROSETTA framework [19, 20].

### F. Step 6: Verify $M_\beta$ with $\mathbb{1}_{W_{\alpha,\beta}} \circ I_\beta$

To verify $M_\beta$ in this last step, we first perform the composition of $\mathbb{1}_{W_{\alpha,\beta}} \circ I_\beta$. This can be intuitively understood as the continued interpretation of the relationship between $W_\alpha$ and $W_\beta$ through the rules of interpretations in $I_\beta$, analogues to how $Q_{\alpha \to \beta} \circ I_\alpha$ in Step 4 was conducted.

The details of $\mathbb{1}_{W_{\alpha,\beta}}$ apparently depends on the type of the relationships between $W_\alpha$ and $W_\beta$. So far, we have only considered the case where $W_\alpha \subset W_\beta$, and lightly commented on $W_\beta \subset W_\alpha$ with respect to reverse engineering. A full account of different types of relationships will be a subject of a future work.

## IV. SYNCHRONIZATION OF MODELS

In this section, we demonstrate how a matrix formalism based on the ROSETTA framework [19, 20] can be utilized to capture the dependencies derived through the core fractal processes between a pair of models, using the ECG case study. Furthermore, we will show, with examples, how such a formalism enables the synchronization of models through structure preserving transformations.

Any graphical model, with textual information suppressed, boils down to a simple graph with nodes and edges, which can then be represented by adjacency or incidence matrices. The ROSETTA framework harnesses the power of such matrix representations to enable: (i) the representation of two models using dedicated adjacency matrices, respectively named as the **N**- and **M**- matrix, and (ii) the representation of the cross-mode dependencies between the two models using an incidence matrix, named as the **Q**-matrix [19, 20].

As illustrated in Figure 6, the Use Case diagram and Activity diagram produced in Section III, illustrated in Figure 3 and 5 respectively, are represented in the **N**- (bottom-left) and **M**- (top-right) matrices respectively in a ROSETTA framework. For each of the matrices, the headers of the rows and columns represent the entities (nodes) in the corresponding diagram. For example, for the Use Case diagram in Figure 3, we have total of five nodes, thus represented by a 5-by-5 matrix with each row/column representing one of these five nodes. Then, the relations between these nodes (edges), also referred to as *in-model dependencies*, are captured in the corresponding matrix element by a statement, $W_{i,j}$, that describes its semantics. For instance, the dependency between $U_1$ and $\mathbb{S}$, is an 'allocation' dependency where the 'Receive Torque Demand' Use Case is allocated to (or owned by) the system, 'ECG'; this is captured in the matrix element in the $U_1$-th row and $\mathbb{S}$-th column, by a statement $W_{U_1,\mathbb{S}}$ that simply states 'allocated to', using a row-to-column convention. It is worth noting that this dependency is not directed, meaning that one can also create a symmetric matrix element to indicate the fact that system ECG owns the Use Case, $U_1$. The rest of the N-matrix also capture other allocations and associations. For simplicity, we will not capture the symmetric counterparts for the undirected edges in the matrix.

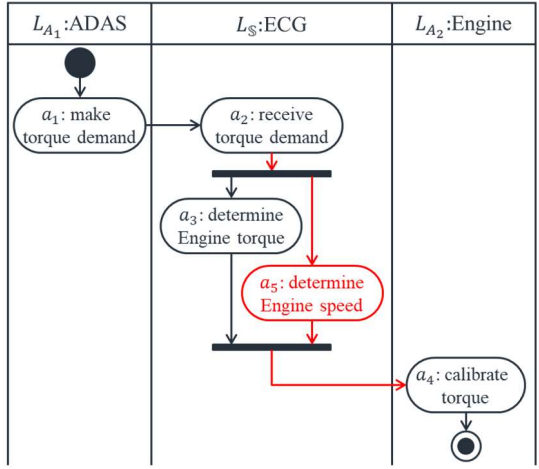

Figure 7. Revised $M_\alpha$ based on the change made in the functional requirements.

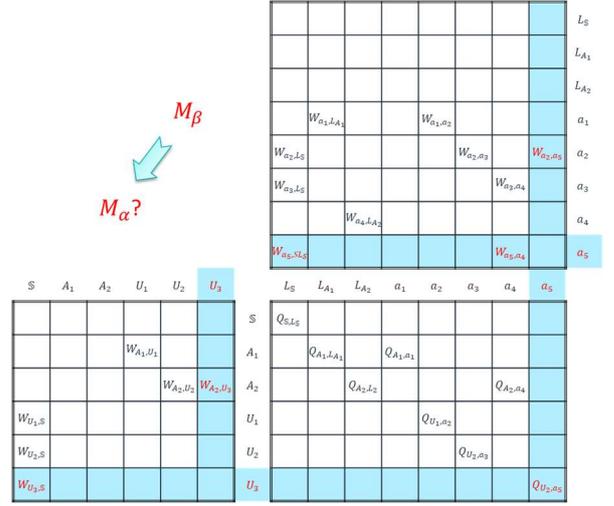

Figure 8. Syncrhonisation of changes using the ROSETTA framework [19,20].

In the **M**-matrix, where the Activity diagram is represented by a 7-by-7 matrix, capturing three Swimlanes and four Actions, and filled with a set of allocation dependencies that reflects the allocations of Actions to Swimlanes, and a set of (directed) precedence dependencies that reflects the flows between the Actions.

Finally, the **Q**-matrix captures the cross-model dependencies that were established through the transformations and the composition of transformations described in the fractal processes. For examples, the System and the Actors are mapped directly into the corresponding Swimlanes through the matrix element, $Q_{\mathbb{S},L_\mathbb{S}}$, $Q_{A_1,L_{A_1}}$ and $Q_{A_2,L_{A_2}}$, respectively, as depicted in Figure 6. The matrix element in the **Q**-matrix notably is about entity-to-entity mappings, whether it is from an Actor to a Swimlane or from a Use Case to a set of Actions. Relation-to-relation mappings are not directly reflected in the **Q**-matrix, but are preserved through the relational transformation [17, 21]:

$$(y_i, y_j) \in \mathbf{N} \text{ with } (y_i, x_k), (y_j, x_l) \in \mathbf{Q} \rightarrow (x_k, x_l) \in \mathbf{M} \quad (1)$$

with $y$ and $x$ representing the model elements, and an ordered pair, e.g., $(y_i, y_j)$ representing the binary relation between the two model elements. Using Figure 6 as an example, tracing through the directed arrow (in green), the allocation relation between the Use Case, $U_1$ and the System, reflected by the matrix elements, $W_{U_1,\mathbb{S}}$, is preserved through the relational transformation into an allocation relation, $W_{a_2,L_\mathbb{S}}$, from the Action, $a_2$ to the Swimlane, $L_\mathbb{S}$. This transformation is mathematically expressed by:

$$(U_1, \mathbb{S}) \in \mathbf{N} \text{ with } (U_1, a_2), (\mathbb{S}, L_\mathbb{S}) \in \mathbf{Q} \rightarrow (a_2, L_\mathbb{S}) \in \mathbf{M} \quad (2)$$

It is worth emphasizing that the relational transformation not only preserves the relationships, but also the intended meaning, i.e., the semantics of that relationship, e.g., allocation in the example above.

The **Q**-matrix together with the relational transforms capture all the cross-model dependencies and confirms their correctness, thereby ensuring the consistency between the two models. Note also that the relational transformation also applies in reverse from the **M**-matrix to the **N**-matrix in a similar way, enabling backward consistency tracing. The two models at this point are said to be *synchronized*. Any change made to any of the two models will likely desynchronize the models. We will show how the model can be re-synchronized by matrix manipulations in the subsequent section.

V. SYNCHRONIZATION OF MODEL CHANGES

To demonstrate the synchronization of changes made to models, we introduce a specific change to the Activity diagram, $M_\beta$. Imagine the scenario where following a discussion with domain engineers who work with combustion engines, it is realized that to achieve a torque demand from the ADAS, one also needs to determine an engine speed value in combination with the engine torque value. As such, the ECG needs to take a further action, 'determine Engine speed' in parallel with the 'determine Engine torque' action. To capture this design decision, the revised Activity diagram in Figure 7 highlights the necessary model changes in red.

To synchronize these changes, we first update the corresponding matrix, the **M**-matrix in Figure 6. This is achieved by first adding a new row and column (shaded in blue) for the new Action, $a_5$, and then adding three new matrix element (highlighted in red) to reflect the following new relationships added to the Activity diagram in Figure 7:

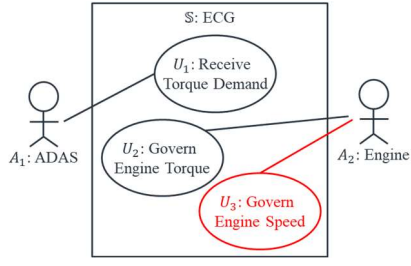

Figure 9. Syncrhonised $M_\beta$.

(i) the allocation of $a_5$ to $L_\mathbb{S}$; (ii) precedence from $a_2$ to $a_5$; and (iii) precedence from $a_5$ to $a_4$. The resultant matrix is depicted in the **M**-matrix in Figure 8.

The addition of $a_5$ to the M-matrix prompts a question: should this be mapped to an existing element in the **N**-matrix through the **Q**-matrix, or a new element to be created? In other words, a decision is to be made by the system engineer on whether the new action can be mapped onto an existing Use Case, or a new dedicated Use Case is necessary. This is how the joint-cognitive nature of the approach comes in.

For the purpose of demonstrating a generalizable pattern, we make the decision that a new Use Case is necessary. This decision results in the creation of a new row/column in the N-matrix, named $U_3$, and a mapping between $U_3$ and $a_5$ in the **Q**-matrix.

Next, we look at new matrix elements added to the M-matrix and how they should be synchronized:

$W_{a_2,a_5}$: precedence from $a_2$ to $a_5$. Applying relational transformation to this relation would result a relation between $U_1$ and $U_3$ with a 'precedence' semantics. However, *precedence between Use Cases is out of the scope* of *a* Use Case diagram as constrained by its metamodel. Therefore, $W_{a_2,a_5}$ would not result in any new relation to be added to the Use Case diagram

$W_{a_5,a_4}$: precedence from $a_5$ to $a_4$. Applying relational transformation to this relation would result a relation between $U_3$ and $A_2$ with a 'precedence' semantics. *This would transform into a* relation between a Use Case and an Actor*, which* is an association*, but* embod*ies specifically the semantics of a* functional flow. Hence, a new matrix element, $W_{A_2,U_1}$, is created in the **N**-matrix. This operation is consistent with the 3$^{rd}$ rule in the Semantic Transformation from Use Case structure to Activity structure presented in Section II-A.

$W_{a_5,L_\mathbb{S}}$: allocation of $a_5$ to $L_\mathbb{S}$. Applying relational transformation to this relation would result a relation between $U_3$ and $\mathbb{S}$ with an 'allocation' semantics. Hence, a new matrix element, $W_{U_3,Sys}$, is created in the **N**-matrix.

The changes made to the **N**-matrix thus far completely synchronizes the changes introduced in the **M**-matrix. We then, convert the **N**-matrix back into a new Use Case diagram, as illustrated in Figure 9, which is a consistent model with the Activity diagram in Figure 7.

In this example, we have shown how *creat*e-type of changes are handled through matrix manipulation. The formalism is also capable of dealing with other basic types of changes including *delete* (e.g., removing an entity) and *update* (e.g., changing the naming an entity), as in the well-known basic operations, CRUD, in computer programming. The degree of automation possible and how manipulation facilitates a joint cognitive design approach is subject to future research.

## VI. VALIDATION USING PRACTICAL METHODOLOGIES

To validate the use of structure preserving transformations as a practical methodology, we show that it is coherent with a representative sample of well-known MBSE Architecture Definition and Systems Engineering methodologies, including OOSEM [3], aMBSE [22] and Arcadia [23, 24]. To facilitate the comparison, we have analyzed the essential Architecture Definition activities within these methodologies and amalgamated them into a generic suite of activities that can be compared with the fractal processed in the core formalism, as depicted in Figure 10.

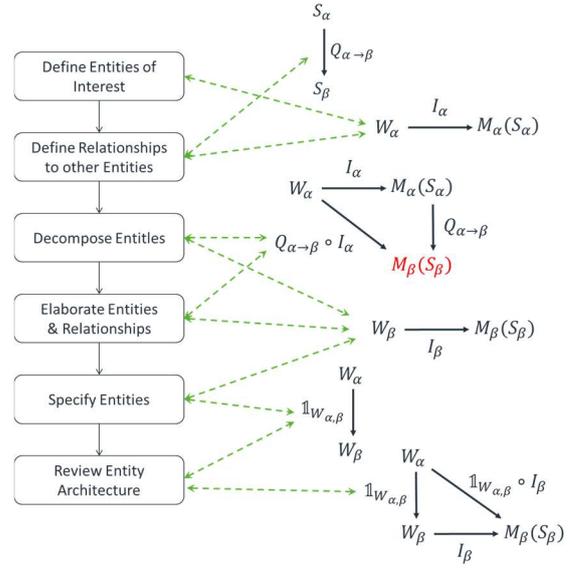

Figure 10. Mapping of generic engineering activities to the fractal processes

Each of the representative methodologies differs in specifics such as the ordering of activities and the details of prescribed analyses but they are all rather similar in their main features. Fundamentally, the methodologies work primarily top-down to link a 'problem space' with a 'solution space'. The problem space is focused firstly on stakeholders and their requirements, usually via analysis of operational activities, and secondly on the derived system requirements, usually by analysis of the stakeholder requirements to derive functional and performance implications for the system. The solution space is then usually focused on logical and physical architecture, establishing the logical and physical elements and associated architectural structures necessary to satisfy the system requirements. Within this basic top-down flow there are generally iterations and revisions necessitating synchronization across models.

We can further abstract the individual activities within the solution space to define a generic process template applicable to any architectural stage as in Figure 10. In essence, the process sequence relates entities of a particular type, which are associated with other entities in a structure, followed by decomposition into entities of the same type in the same type of structure, then elaboration using domain knowledge into

entities of different types in different types of structure. In detail, this follows the sequence pattern:

- Define entities of interest
- Define relationships to other entities
- Decompose entities and relationships
- Elaborate entities and relationships
- Specify entities in detail
- Review entity architecture

The entities could be, say, logical components or physical components.

It is then possible to map this generic template to the core fractal processes identified in this paper, as shown in Figure 9. Hence, there is good coherence between a generic informal process capturing several well-known MBSE process variants and between our approach.

In practice, application of systems engineering can be more complicated than implied by this mapping. We have restricted the generic sequence to what is essentially a top-down paradigm, when other tactics can be equally important, including bottom-up, middle-out and outside-in [5, 25]. In essence, this is a matter of traversing the basic generic process framework in a different order, with different starting points and with different relationships between $W_\alpha$ and $W_\beta$. For example, a bottom-up tactic would utilize the concept of reversing engineering and abstraction, as explained in Section III-E. Whatever the model change, the essential synchronization ensures coherence of related elements and structures reached via different trajectories. In complex systems engineering, the need to synchronize different tactics employed concurrently is commonplace. The core formalism enables this by supporting reverse transformations and explicit synchronization where appropriate.

## VII. Conclusion

In summary, this work presents a theoretical formalism underpinned by structure preserving transformations to achieve consistent development and specification of models; and thereby enables the synchronization of models in ways that are agnostic of where the changes to models have been introduced. The fractal processes break the formalism down to a series of elementary steps, enabling the realization of the formalism in a joint cognitive approach, with automated model development and synchronization to the maximum extent possible and support to system engineers making informed decisions where automation is not generalizable or impractical.

Through the demonstrated application of the fractal processes and synchronization using a practical case study on a ADAS ECG system, and discussion around practical MBSE methodologies, this work has addressed the challenge in practical model-based systems engineering where models are often developed in ways that are not resilient to changes and thus not reusable, making them no better than documents. Further, the case study focused on the synchronization of models conforming to the same metamodel (UML/SysML); hence, a specific research direction to be explored in a future work is to investigate the synchronization of models that are of different types, e.g., between a UML/SysML model and a Fault Tree [26].